\documentclass[pdflatex,sn-mathphys-num]{sn-jnl}

\usepackage{graphicx}
\usepackage{multirow}
\usepackage{amsmath,amssymb,amsfonts}
\usepackage{amsthm}
\usepackage{mathrsfs}
\usepackage[title]{appendix}
\usepackage{xcolor}
\usepackage{textcomp}
\usepackage{manyfoot}
\usepackage{booktabs}
\usepackage{array}
\usepackage{longtable}
\usepackage{bookmark}
\hypersetup{hypertexnames=false}

\theoremstyle{thmstyleone}

\theoremstyle{thmstylethree}

\begin{document}

\title[System Attribution in LLM Brand Recommendations]{System Attribution in LLM Brand Recommendations: Single Responses Identify the System, Aggregated Brand Profiles Do Not Transfer}

\author*[1,2]{\fnm{Dmitrij} \sur{\.Zatuchin}}\email{dmitrij.zatuchin@eek.ee}

\affil*[1]{\orgdiv{Department of Information Technologies},
  \orgname{Estonian Entrepreneurship University of Applied Sciences (EUAS)},
  \orgaddress{\city{Tallinn}, \country{Estonia}}}
\affil[2]{\orgname{Rankfor.AI},
  \orgaddress{\city{Wroc\l{}aw}, \country{Poland}; \city{Tallinn}, \country{Estonia}}}

\abstract{Audits of AI visibility summarise the brand recommendations of deployed language models into per-system profiles. This paper asks whether such a profile describes the system, and puts the question to one corpus at two units of observation. The corpus holds 6{,}475 stored responses, 6{,}324 of them analysable, collected between December 2025 and February 2026 from five deployed endpoints across gift-recommendation, corporate-reputation and category-ownership queries. The collection harness cut many answers short: 83.1\% of the Gemini~3 Flash answers in category ownership end mid-sentence under a 1{,}024-token output cap. With every answer cut to its first 800 characters, so that where an answer stops carries no information, a character $n$-gram classifier cross-validated by prompt attributes a single response to GPT-5.2, Gemini~3 Flash, Gemini~3 Flash with search, Grok or Perplexity sonar-pro with 97.84\% accuracy (5{,}028 responses, 383 prompts, majority class 31.5\%, 30 split seeds); full-length text gives 98.14\%. Length alone then falls to the majority rate, 24 formatting statistics reach 95.79\%, and masking brand names and capitalised tokens leaves 97.72\%. Held-out query conditions inside the same harness keep 97.43\% weighted by size and 88.0\% unweighted; in the one held-out condition that changes the harness, a retrieval-grounded arm, no Grok answer is attributed to Grok (0/120). Aggregated into 50 model-by-domain-by-condition units, twelve behavioural features separate four systems at 66.53\% under grouped cross-validation, against a label-permutation null with mean 33.71\% and 95th percentile 46.0\%; two of the four systems occur in one domain only, and the feature carrying the separation tracks domain and repetition count. Across domains the aggregate profile fails: a forest trained on ten category-ownership units assigns all 22 gift units to the wrong system, consistent with a reversal in brand volume between the two systems (8.41 against 0.94 brands per response in gifts, 3.01 against 3.91 in category ownership), while single responses transfer at 89.92\% balanced accuracy. The surface form of an answer carries the system across the query domains tested; aggregated brand behaviour does not, and the uncrossed design cannot separate the system from the domain or the harness.}


\maketitle

\section{Introduction}\label{sec:intro}

Large language models now answer questions that used to go to a search engine: what to buy, which supplier to shortlist, what a company's reputation is~\cite{brown2020gpt3, davenport2020ai}. Generative engine optimisation treats those answers as a surface that content can be shaped for and measured on~\cite{aggarwal2024geo}. This paper starts from that premise and tests the assumption beneath any per-system measurement of it: that an answer can be attributed to the system that produced it, and that what the measurement observes is a property of that system. The query domain and the collection harness are the two competing sources this corpus lets us examine.

The test uses one corpus at two units of observation, and the two units give different answers.

The first unit is the single stored response. A character $n$-gram classifier identifies which of five deployed endpoints produced an answer with 97.84\% accuracy under cross-validation that holds out whole prompts, when every answer is cut to its first 800 characters. The cut matters because the harness truncated one system's answers far more often than the others', and a classifier can read truncation. Layout carries most of the signal: 24 formatting statistics reach 95.79\% on the same prefixes.

The second unit is the model-by-domain-by-condition aggregate, the unit an AI-visibility report summarises. Twelve behavioural features of brand volume, concentration, sentiment and source behaviour separate four systems at 66.53\% inside the collection window, and a classifier moved to a different query domain assigns every gift-domain unit to the wrong system. The brand-volume ordering of the two systems present in both domains reverses between them.

The discriminative information about a deployed endpoint that survives a change of query domain in this corpus sits in the surface form of its answers. The brand behaviour that practitioners summarise, and that the earlier version of this work treated as a system signature, varies with the query domain, and the design cannot say how much of it belongs to the system.

We state four questions.

\begin{description}
    \item[\textbf{RQ1}] Given a single stored response to a brand-related prompt, how accurately can the deployed endpoint that produced it be identified, and how well calibrated is that identification?
    \item[\textbf{RQ2}] What carries the response-level signal: which brands the answer names, or how the answer is laid out?
    \item[\textbf{RQ3}] How much of the signal survives a change of query domain, at each of the two units of observation?
    \item[\textbf{RQ4}] What do aggregated behavioural profiles add, and where do they fail?
\end{description}

The contributions are four. A response-level attribution result on consumer-style brand queries collected without crafted probes, with a truncation control, calibration measured inside and outside the training distribution, and held-out collection conditions. A measurement negative: aggregated brand-behaviour profiles that separate systems inside one domain and invert across domains, with the reversal quantified on complete answers. A corpus description that states its collection defects: truncation in three subsets, empty responses concentrated on long-answer prompts, completion caps that differ between subsets for one endpoint, and one arm whose prompts ask for sources. And Appendix~\ref{secD}, which lists the values of the earlier version of this work that do not reproduce from the raw files.

Appendix~\ref{secD} lists every value that changed from an earlier, unpublished version of this analysis. Every number here comes from a rerun of the analysis against the raw response files.

\section{Related Work}\label{sec:related}

\subsection{Which Model Wrote This}

Fingerprinting a language model from its outputs has an active line and a passive line. On the active side, LLMmap~\cite{pasquini2024llmmap} identifies 42 model versions from eight crafted interactions at over 95\% accuracy, Iourovitski et al.~\cite{iourovitski2024hideseek} search prompts evolutionarily for family-level identification, and UTF~\cite{cai2024utf} reads under-trained tokenizer artefacts. Adi et al.~\cite{adi2018turning} and Cao et al.~\cite{cao2021ipguard} set the precedent in the wider deep-learning security literature. These methods choose their own inputs. The present setting cannot: the prompts are consumer-style questions collected for other purposes, and the only evidence is the stored answer.

On the passive side, McGovern et al.~\cite{mcgovern2025fingerprints} show that simple classifiers over character, word and part-of-speech $n$-grams separate the outputs of different models and model families, and that these fingerprints hold across text domains. Sun et al.~\cite{sun2025idiosyncrasies} fine-tune text embedding models to predict which of five deployed chat systems produced a text, reporting 97.1\% five-way accuracy; a TF-IDF logistic regression over word unigrams and bigrams reaches 85.5\% on the same chat-system task, word-shuffled text keeps 88.9\%, and markdown elements alone give 73.1\%. Bitton et al.~\cite{bitton2025stylistic} attribute text to model families from stylistic features. Stylometric attribution has since reached short samples~\cite{przystalski2026stylometry}, book-length text~\cite{shetty2026ghostwriters} and source code~\cite{bisztray2025codestylometry}. The response-level result here is consistent with this body of work and uses its methods. The increments are the application domain, brand recommendation; two retrieving endpoints among the five; a truncation control; calibration measured under held-out conditions and across domains; held-out collection conditions, including one that changes the harness; and the contrast with the aggregated unit that AI-visibility reports summarise.

Detection of machine authorship is a different, binary task~\cite{mitchell2023detectgpt, kirchenbauer2023watermark, gehrmann2019gltr, hans2024binoculars}, surveyed through 2025 by Wu et al.~\cite{wu2025survey}. Authorship attribution over human writers is the older discipline that supplies our model: Stamatatos~\cite{stamatatos2009survey} surveys it, and Sapkota et al.~\cite{sapkota2015notall} show which character $n$-grams carry attribution signal and which carry topic.

\subsection{Behaviour of Deployed Systems and Its Drift}

Chen et al.~\cite{chen2024chatgpt} document behaviour shifts in a single endpoint across calendar months without any architectural change. Ouyang et al.~\cite{ouyang2022training} describe the alignment machinery that produces such shifts. Kamruzzaman et al.~\cite{kamruzzaman2024brand} report systematic brand bias in LLM recommendations. \.Zatuchin~\cite{zatuchin2026a} reports the inter-system differences in gift-recommendation brand sets that part of the present corpus was collected for, under the repetition protocol of~\cite{zatuchin2026c}. Hicks et al.~\cite{hicks2024chatgpt} argue that LLM outputs are better assessed as statistical behaviour than as truth claims, which is the posture of the measurement here.

\subsection{Retrieval-Augmented Services Are Pipelines}

Lewis et al.~\cite{lewis2020rag} formalised retrieval-augmented generation; Liu et al.~\cite{liu2023evaluating} evaluated verifiability in commercial generative search engines; RAGAS~\cite{es2024ragas} and RGB~\cite{chen2024rgb} supply RAG-specific evaluation. Two of the five endpoints in this corpus retrieve: Perplexity sonar-pro, and Gemini~3 Flash with the search tool attached. A signal that separates them from a parametric endpoint may come from the retrieval layer, the citation policy or the re-ranking stage, so this paper labels those results at the level of the system.

\subsection{Small Samples, Grouped Data, Calibration}

Random forests~\cite{breiman2001random} and principal component analysis~\cite{jolliffe2002pca} carry the aggregated analysis; at 50 units their outputs are descriptive. Hurlbert~\cite{hurlbert1984pseudo} is the reference for pseudo-replication, the central design problem of the aggregated unit; Varoquaux~\cite{varoquaux2018cross} shows how wide cross-validation error bars become at small $n$; Kaufman et al.~\cite{kaufman2012leakage} catalogue the leakage that grouped cross-validation is meant to prevent. Permutation tests~\cite{good2005permutation} and the bootstrap~\cite{efron1979bootstrap} replace parametric tests, and calibration is measured with the binned expected calibration error of Naeini et al.~\cite{naeini2015obtaining} in the form used by Guo et al.~\cite{guo2017calibration}. All classifiers are scikit-learn implementations~\cite{pedregosa2011scikit}.

\section{Corpus}\label{sec:data}

\subsection{What Was Collected}

The corpus is the union of three query domains collected for earlier studies between December 2025 and February 2026 and re-read here from the raw response files. Every file in the source directories is read and every row is accounted for: the loader writes a per-file manifest with row counts and md5 digests and asserts the per-corpus totals. The count is 121 files and 6{,}475 stored responses.

Two classes of row are set aside before any feature is computed.

\paragraph{Empty responses.} 136 stored GPT-5.2 responses in the category-ownership subset have zero-length text and no error flag. They are a behaviour of that endpoint under the 1{,}024-token completion cap of that collection, and they are missing not at random. They fall on 43 of the 250 prompts, 13 of them empty on every repetition, and on long-answer prompts: the GPT-5.2 answers that survive on those prompts have a median of 4{,}104 characters against 2{,}654 elsewhere. By industry they number 64 in healthtech, 33 in fintech, 22 in SaaS, 14 in consulting and 3 in e-commerce, so the GPT-5.2 healthtech unit is built from 186 of 250 responses. They are excluded from every feature by default, and Section~\ref{sec:empties} reports two sensitivity policies. The earlier version of this work counted them as answers naming no brand.

\paragraph{Rows with a colliding iteration index.} 15 GPT-5.2 rows in the parametric reputation subset share a (file, system, prompt, iteration) key with an earlier row. None of the 15 texts matches the retained row: they are distinct generations stored under a reused iteration number. They are dropped by keeping the first row per key. The subset is outside the response-level headline, and keeping them moves the one aggregate unit they belong to by 0.010 in mean brand count and 0.011 in abstinence.

Every other row is analysable: 6{,}324 responses remain. A gate in the earlier pipeline that tested the parsed JSON for a dictionary type silently skipped 72 files, dropping 360 responses of the grounded reputation arm. Those rows are present here.

Table~\ref{tab:corpus} describes the corpus as the files record it.

\begin{table}[t]
\caption{The corpus as read from the raw files. Responses are the analysable rows, after the 136 empty and 15 colliding-index rows are set aside. Incomplete is the share of each system's responses whose text stops before a finished sentence (Section~\ref{sec:truncation} gives the test), listed per system in the order of the Systems column.}\label{tab:corpus}
\centering\footnotesize\setlength{\tabcolsep}{3pt}
\begin{tabular*}{\textwidth}{@{\extracolsep\fill}llrrrl@{}}
\toprule
\textbf{Subset} & \textbf{Systems} & \textbf{Resp.} & \textbf{Prompts} & \textbf{Cond.} & \textbf{Incomplete (\%)} \\
\midrule
Gifts, Christmas       & Gem., GPT, Grok            &   904 &   8 & 6 & 9.2 / 3.5 / 0.0 \\
Gifts, Valentine's     & Gem., GPT, Grok            &   150 &   5 & 5 & 0.0 / 0.0 / 0.0 \\
Reputation, grounded   & Gem.-search, Grok, Perpl.  &   360 & 120 & 1 & 0.8 / 0.0 / 0.0 \\
Reputation, parametric & Gem., GPT                  & 1{,}296 & 360 & 1 & 70.8 / 2.2 \\
Category ownership     & Gem., GPT, Perpl.          & 3{,}614 & 250 & 5 & 83.1 / 2.2 / 0.5 \\
\midrule
\textbf{Total}         & five endpoints             & \textbf{6{,}324} & \textbf{743} & \textbf{18} & \\
\bottomrule
\end{tabular*}
\end{table}

The five endpoints are GPT-5.2 (OpenAI, chat completions), Gemini~3 Flash (Google, GenerateContent), Gemini~3 Flash with the \texttt{google\_search} tool attached, Grok (xAI), and Perplexity sonar-pro. Gemini with search is kept as a separate label: it is a different deployment of the same base model, and merging it into the parametric Gemini label would hide the retrieval layer inside a system identity.

\subsection{What the Collection Protocol Actually Was}\label{sec:protocol}

The collection scripts and the stored rows disagree with the earlier text on four points, and the files are taken as authoritative.

Repetitions per system and prompt, as the stored rows record them, are 40 in the Christmas gift subset, 10 in the Valentine's subset, 5 in category ownership and 1 for most reputation cells. The claim of five repetitions for every prompt holds for the category-ownership subset and for three reputation templates.

Sampling temperature 0.3 is recorded in the collection code of the reputation and category-ownership subsets. The gift subset stores no sampling parameters.

The system version strings recorded are \texttt{gpt-5.2}, \texttt{gemini-3-flash-preview} and \texttt{sonar-pro}. No version string is recorded for the Grok rows of the gift subset; the only Grok version string in any collection script is \texttt{grok-3}, in the grounded arm. The label Grok may therefore pool two Grok versions, and this paper names the provider and states that the gift-subset version was not logged.

Per-row timestamps run from 2025-12-23 to 2026-01-05 (Christmas gifts), 2026-02-15 (Valentine's), 2026-02-20 (both reputation subsets) and 2026-02-23 to 2026-02-24 (category ownership).

\subsection{Collection Confounds, Stated Before the Results}\label{sec:truncation}

A classifier that reads the surface of an answer reads the harness that produced it. Four differences between the arms of this corpus are properties of the collection.

\paragraph{Answers are cut short in three subsets.} A response counts as complete when, after trailing Markdown emphasis and whitespace are stripped, its last character closes a sentence, a bracket, a quotation or a table row, or is an emoji. By that test 83.1\% of the Gemini~3 Flash answers in category ownership stop mid-sentence, against 2.2\% for GPT-5.2 and 0.5\% for Perplexity; the collection capped every endpoint at 1{,}024 output tokens. In the parametric reputation subset the Gemini figure is 70.8\%, with a median stored length of 104 characters against 1{,}894 for GPT-5.2, which is a storage defect; in the Christmas gift subset it is 9.2\%. A stricter test that also counts every answer ending in closing Markdown or a colon as incomplete gives 88.2\%, 71.8\% and 39.4\% for the same three cells. Of the 5{,}028 responses in the headline task, 1{,}107 are incomplete, 1{,}065 of them Gemini~3 Flash.

Two measurements answer this. Every response-level task is run three ways: on the stored text; with every text cut to its first 800 characters, which removes the ending from all but 24 responses and with it the information in where an answer stops; and on complete responses only. The parametric reputation subset, whose truncation also removes most of the content of the Gemini arm, is excluded from the headline and reported descriptively in Section~\ref{sec:reputation-descriptive}.

\paragraph{Completion caps differ between subsets.} The category-ownership collection capped GPT-5.2 at 1{,}024 completion tokens; the reputation collection used 16{,}384 for the same endpoint. The empty GPT-5.2 responses of Section~\ref{sec:data} and the truncated Gemini answers above are two faces of the same cap.

\paragraph{One arm was asked for its sources.} The grounded reputation arm was collected with web search enabled for all three of its endpoints, and three of its five prompt templates ask the system to name its sources. No other subset does either. Citation-related formatting therefore carries information about the arm as well as the endpoint.

\paragraph{Repetition count and temperature differ by subset.} Any feature computed across the repetitions of a prompt depends on how many repetitions there were (40, 10, 5 or 1) and at what temperature.

\section{Method}\label{sec:methods}

\subsection{Two Units of Observation}\label{sec:units}

The response unit is one stored answer. Its label is the endpoint that produced it. There are 6{,}324 of them, and 5{,}028 once the parametric reputation subset is excluded.

The aggregate unit is a model-by-domain-by-condition cell: one system, one query domain, one collection condition such as a recipient context or an industry. There are 50 such units over four systems, and this is the unit the earlier version of this work classified. A unit summarises between 10 and 648 responses, median 40, and every unit enters the classifier with equal weight.

The grounded Gemini arm is a fifth label at the response level, with 120 responses in 120 prompt groups. At the aggregate level it is a single unit in a single condition, so grouped cross-validation holds out its only example by construction; it is described in Section~\ref{sec:grounded} and not classified. The aggregate task has four labels. Two of them are tied to a domain: Grok units exist only in the gift domain and Perplexity units only in category ownership.

\subsection{Response-Level Model}\label{sec:text-model}

Text is represented as character $n$-grams within word boundaries, $n$ from 2 to 4, TF-IDF weighted with sublinear term frequency, minimum document frequency 3, at most 150{,}000 features, and classified with multinomial logistic regression at $C = 10$~\cite{pedregosa2011scikit}. Character $n$-grams are the standard representation for authorship attribution~\cite{stamatatos2009survey} and they capture punctuation, list markers and casing along with words.

The grouping unit for cross-validation is the prompt. No prompt appears in both a training fold and a test fold. Accuracy is the mean over 30 split seeds of stratified grouped 5-fold cross-validation, with the standard deviation across seeds. The logistic regression is deterministic given the split, so that standard deviation is split variance only.

\subsection{Behavioural Feature Vector}\label{sec:features}

Each aggregate unit is a vector in $\mathbb{R}^{12}$, in four families. Where the earlier code computed something else, Appendix~\ref{secD} records the difference.

\paragraph{Brand volume.} Mean brand count $\mu_b$, the average number of \emph{unique} canonical brands per response, identified by a deterministic dictionary pass over a per-domain vocabulary. Brand-count coefficient of variation $\mathrm{CV}_b = \sigma_b / \mu_b$ over responses in the unit. Brand-abstinence rate $\alpha$, the proportion of non-empty responses with no detected brand.

\paragraph{Concentration.} Gini coefficient $G$~\cite{gini1912variabilita} over the brand-frequency distribution of the unit. Normalised Shannon entropy $H_{\mathrm{norm}} = H / \log_2 N$ over the $N$ unique brands of the unit~\cite{shannon1948}. Top-brand share $\tau$.

\paragraph{Sentiment and tone.} Mean VADER compound score $\bar{s}$ over the non-empty responses of the unit~\cite{hutto2014vader}, its variance $\sigma_s^2$, and positivity rate $\pi$. On this corpus these three features track response length (Section~\ref{sec:limitations}), and Section~\ref{sec:unit-results} reports the vector without them.

\paragraph{Behaviour.} Response consistency $\bar{c}$, the mean pairwise TF-IDF cosine similarity across the repetitions of a prompt, averaged over the prompts of the unit. Citation-absence rate, the proportion of responses whose text contains no URL or bracket citation. Cross-query brand diversity $D_q$, unique brands across the unit divided by its number of queries.

Two of the twelve are close to constant. Citation absence is exactly 1.0 in 37 of the 50 units, and positivity rate is exactly 1.0 in 40 of them.

The brand dictionary was checked on 50 randomly drawn GPT-5.2 gift answers against a reference built by the analysis agent, one automated pass with no human rater. Precision was 0.827 (9 of 52 hits were ordinary words such as ``beats'' and ``block''), and recall at most 0.43: the reference finds 2.0 brands per answer against the dictionary's 1.04, mostly games, toys and platforms outside the vocabulary. All 18 answers in which the dictionary found no brand also had none in the reference.

The source-opacity feature of the earlier version, the share of source attributions typed as unattributable, is defined only where an endpoint emits a structured attribution payload. Only the grounded arm does, so the feature is left out of the vector and reported in Section~\ref{sec:grounded}.

\subsection{Aggregate-Level Model and Protocol}\label{sec:unit-model}

A random forest of 500 trees, unconstrained depth, minimum samples per split 2, minimum samples per leaf 1, class-balanced weighting, with features standardised inside the cross-validation pipeline~\cite{breiman2001random}. The grouping unit is the (domain, condition) pair, so two units from the same recipient context or the same industry never straddle a fold boundary~\cite{kaufman2012leakage}. Accuracy is the mean over 30 split seeds of stratified grouped 5-fold cross-validation. The forest's own seed is fixed at 42 for the headline, so its standard deviation is split variance; Section~\ref{sec:unit-results} also reports the figure with the forest seed varied alongside the split seed. Leave-one-group-out is reported as the stricter check, and an RBF support vector machine as a baseline. The ungrouped protocol is also run, over 100 split seeds, because the earlier version reported a number that came from it while describing a grouped protocol.

\subsection{Uncertainty and Calibration}\label{sec:stats}

Permutation nulls shuffle the labels within groups, so the null keeps the cluster structure and the tie between two labels and one domain: 1{,}000 shuffles at the aggregate level, with $p = (1 + |\{i : \mathrm{acc}_i \geq \mathrm{acc}_{\mathrm{obs}}\}|) / (1 + n_{\mathrm{perm}})$ computed for the statistic at split seed 0. At the response level 200 refits of the primary model are not affordable, so the null uses a lighter surrogate, a linear support vector machine over character 2- and 3-grams, with its own observed accuracy reported; 200 shuffles bound $p$ at 0.005. Bootstrap intervals resample the grouping unit, 1{,}000 times, over fixed out-of-fold predictions; they describe test-set sampling conditional on the fitted models and leave out training variance~\cite{varoquaux2018cross}. Calibration is the 10-bin expected calibration error (ECE) of the top-class probability~\cite{naeini2015obtaining, guo2017calibration}, measured on out-of-fold predictions inside the training distribution, on held-out conditions and across domains.

\subsection{Controls}\label{sec:controls-method}

All controls run under the grouped protocol over 30 split seeds, on each of the three text variants.

\emph{Length only}: one feature per response, its character count, classified with logistic regression, a random forest and gradient boosting, so the control does not depend on a linear decision rule.

\emph{Twenty-four formatting statistics}: list-marker share, numbered-line share, heading share, bold rate, dash rates, colon, semicolon, question and exclamation rates, parenthesis rate, URL rate, bracket-citation rate, pipe rate, emoji rate, mean word length, type-token ratio, uppercase share, digit share, mean sentence length, characters, words and lines, with a random forest. None of these sees a word's identity; type-token ratio and mean word length are lexical statistics, and uppercase and digit share partly track proper nouns.

\emph{Masking}: every span matched by the case-insensitive brand dictionary, every token whose first letter is an uppercase letter in any script, every URL and every digit run replaced before vectorisation.

\emph{Leave-one-condition-out}: a whole collection condition, such as one seasonal recipient or one industry, held out of training, with the confusion and calibration of each held-out condition recorded.

\subsection{Cross-Domain Transfer}\label{sec:transfer-method}

Transfer trains in one domain and tests in another. Gemini~3 Flash and GPT-5.2 are the only systems present in more than one domain, so the transfer task is two-class and chance is 50\%. At the aggregate level four train and test pairs are run with the six brand-volume and concentration features defined in both domains, and with all twelve for comparison. At the response level the gift and category-ownership domains are run in both directions with the classifier of Section~\ref{sec:text-model}, reported as balanced accuracy because label counts differ between domains.

\section{Results}\label{sec:results}

\subsection{Attribution at the Response Level}\label{sec:response-results}

A single stored response identifies its endpoint, and truncation does not produce the result. On the 5{,}028 responses of the headline task, grouped by prompt, five-way accuracy is 98.14\% on the stored text (SD 0.09 over 30 split seeds), 97.84\% with every text cut to its first 800 characters (SD 0.21, prompt bootstrap [96.11, 99.3]), and 97.86\% on the 3{,}921 complete responses alone, against a majority baseline of 31.5\%. Macro F1 on the prefixes is 0.967. The within-prompt permutation null, run with the surrogate model, has mean 33.43\% and maximum 35.18\% against an observed 98.05\% for the surrogate itself, so $p \leq 0.005$.

Table~\ref{tab:response} reports each domain on the 800-character prefixes. The prefix cut costs the gift domain 1.7 points and changes no other domain by more than 0.1.

\begin{table}[t]
\caption{Response-level attribution, mean of 30 split seeds of stratified grouped 5-fold cross-validation with the prompt as the group. Prefix: every text cut to its first 800 characters. Stored: the full stored text. The parametric reputation row is excluded from the headline because its Gemini arm is truncated in storage to a median of 104 characters.}\label{tab:response}
\centering\scriptsize\setlength{\tabcolsep}{3pt}
\begin{tabular*}{\textwidth}{@{\extracolsep\fill}lrrrrrr@{}}
\toprule
\textbf{Task} & \textbf{Resp.} & \textbf{Groups} & \textbf{Majority} & \textbf{Prefix} & \textbf{SD} & \textbf{Stored} \\
\midrule
Gifts, 3 systems              & 1{,}054 &  13 & 34.2\% & 92.37\% & 0.24 & 94.07\% \\
Reputation grounded, 3 systems &   360 & 120 & 33.3\% & 99.91\% & 0.17 & 99.86\% \\
Category ownership, 3 systems & 3{,}614 & 250 & 34.6\% & 100.0\% & 0.0 & 99.97\% \\
\midrule
\textbf{Pooled, 5 systems}    & \textbf{5{,}028} & \textbf{383} & \textbf{31.5\%} & \textbf{97.84\%} & \textbf{0.21} & \textbf{98.14\%} \\
Pooled, complete responses only & 3{,}921 & 383 & 36.7\% & & & 97.86\% \\
\midrule
Reputation parametric, 2 systems & 1{,}296 & 360 & 50.0\% & 98.55\% & 0.28 & 98.52\% \\
\bottomrule
\end{tabular*}
\end{table}

\paragraph{Held-out conditions.} Table~\ref{tab:loco} holds out one collection condition at a time. The sixteen conditions whose labels all appear in training are recipient contexts and industries collected by the same scripts as the training data, so they vary the query and keep the harness. Over those sixteen, accuracy on the prefixes is 97.43\% weighted by condition size and 88.0\% unweighted. The five industries of category ownership score 100\% and supply 3{,}614 of the 4{,}668 held-out responses; the five Valentine's conditions, 30 responses each, score between 60.0\% and 76.67\% against a three-class baseline of 33\%.

The seventeenth condition, the grounded reputation arm, is the one hold-out in the corpus that changes the harness: web search on, prompts that ask for sources, a different script. Its Gemini-with-search label never appears in training, so its accuracy is undefined. For the other two systems it is the only cross-harness test the corpus permits. Perplexity answers are attributed to Perplexity (120/120 on the stored text), and no Grok answer is attributed to Grok (0/120; on the stored text 90 of them are called Gemini~3 Flash, on prefixes 77 are called Perplexity). Most of the Valentine's errors are the same confusion inside the gift domain: Grok answers attributed to GPT-5.2. Endpoint attribution across harnesses fails for Grok in this test. The gift-subset Grok version was not logged and the grounded arm used \texttt{grok-3}, so the failure may be a version difference as much as a harness difference.

\begin{table}[t]
\caption{Leave-one-condition-out at the response level: each collection condition held out of training in turn, accuracy (\%) on the stored text and on 800-character prefixes, and correct attributions per system on the prefixes; in the grounded row the Gem.\ column is Gemini with search (Val.\ Valentine's gifts, Xmas Christmas gifts, Rep.\ reputation, Cat.\ category ownership). $^\dagger$The grounded arm is the only held-out condition that changes the harness; its Gemini-with-search label is absent from training, so its accuracy is undefined.}\label{tab:loco}
\centering\scriptsize\setlength{\tabcolsep}{3pt}
\begin{tabular}{@{}lrrrrrrr@{}}
\toprule
\textbf{Condition} & \textbf{Resp.} & \textbf{Stored} & \textbf{Prefix} & \textbf{Gem.} & \textbf{GPT} & \textbf{Grok} & \textbf{Perpl.} \\
\midrule
Val. boyfriend & 30 & 63.33 & 63.33 & 10/10 & 7/10 & 2/10 & -- \\
Val. girlfriend & 30 & 70.0 & 76.67 & 10/10 & 8/10 & 5/10 & -- \\
Val. husband & 30 & 63.33 & 60.0 & 10/10 & 8/10 & 0/10 & -- \\
Val. partner & 30 & 70.0 & 60.0 & 10/10 & 4/10 & 4/10 & -- \\
Val. wife & 30 & 50.0 & 76.67 & 10/10 & 9/10 & 4/10 & -- \\
Xmas daughter & 120 & 100.0 & 100.0 & 40/40 & 40/40 & 40/40 & -- \\
Xmas generic & 285 & 96.14 & 90.53 & 95/95 & 100/100 & 63/90 & -- \\
Xmas husband & 230 & 83.91 & 80.87 & 41/60 & 55/80 & 90/90 & -- \\
Xmas partner & 30 & 100.0 & 100.0 & 10/10 & 10/10 & 10/10 & -- \\
Xmas son & 120 & 100.0 & 100.0 & 40/40 & 40/40 & 40/40 & -- \\
Xmas wife & 119 & 100.0 & 100.0 & 39/39 & 40/40 & 40/40 & -- \\
Rep. grounded$^\dagger$ & 360 & undefined & undefined & 0/120 & -- & 0/120 & 120/120 \\
Cat. consulting & 736 & 100.0 & 100.0 & 250/250 & 236/236 & -- & 250/250 \\
Cat. ecommerce & 747 & 100.0 & 100.0 & 250/250 & 247/247 & -- & 250/250 \\
Cat. fintech & 717 & 100.0 & 100.0 & 250/250 & 217/217 & -- & 250/250 \\
Cat. healthtech & 686 & 100.0 & 100.0 & 250/250 & 186/186 & -- & 250/250 \\
Cat. saas & 728 & 100.0 & 100.0 & 250/250 & 228/228 & -- & 250/250 \\
\midrule
\multicolumn{2}{@{}l}{Sixteen defined, weighted} & 97.79 & 97.43 & & & & \\
\multicolumn{2}{@{}l}{Sixteen defined, unweighted} & 87.29 & 88.0 & & & & \\
\bottomrule
\end{tabular}
\end{table}

\begin{figure}[ht]
\centering
\includegraphics[width=0.92\textwidth]{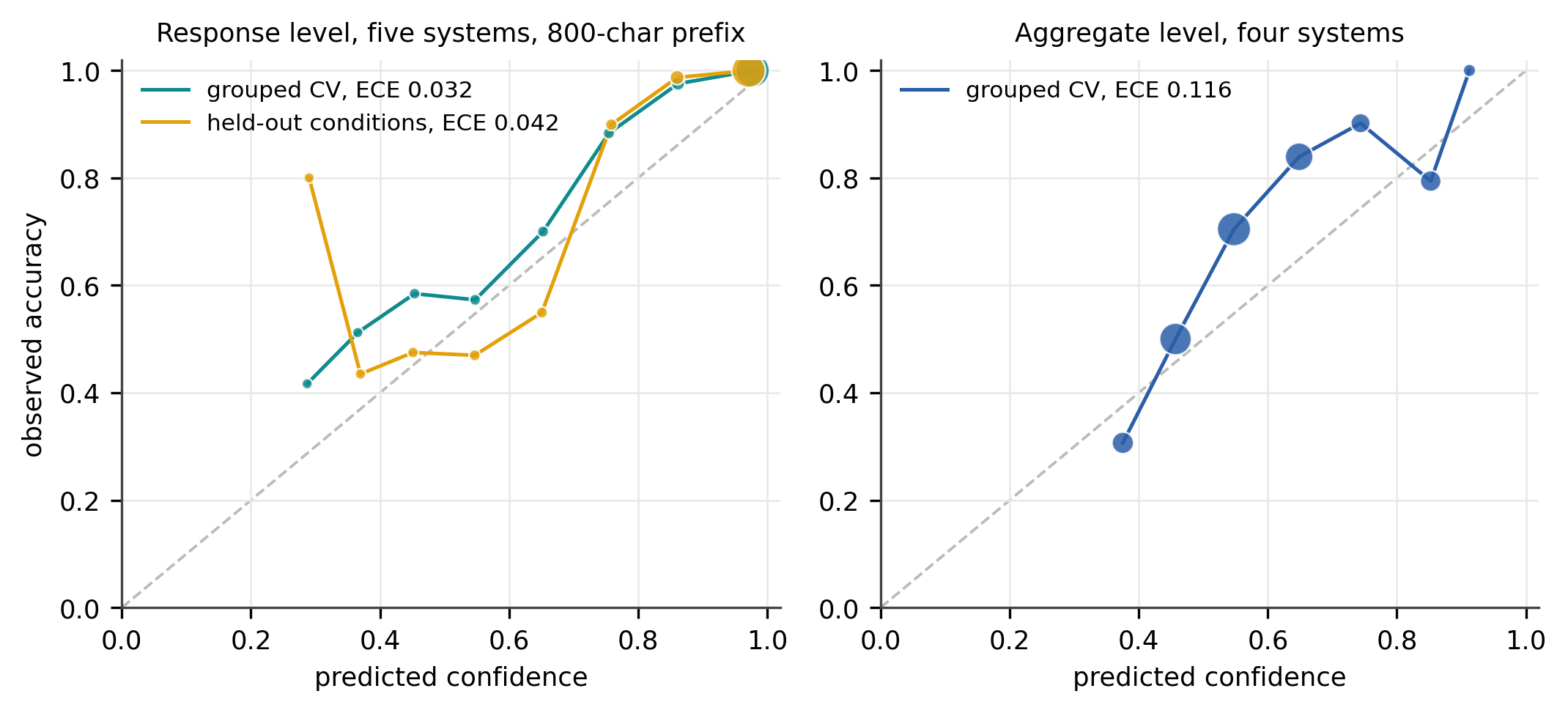}
\caption{Reliability of the two classifiers. Left, the response-level character $n$-gram model on 800-character prefixes, inside the training distribution (grouped cross-validation) and on held-out conditions. Right, the aggregate-level random forest under grouped cross-validation. Points on the diagonal mean that a stated confidence matches the observed accuracy at that confidence. Marker area counts predictions in the bin, and the grouped curves pool the predictions of 30 split seeds, so each response or unit contributes 30 times.}
\label{fig:calibration}
\end{figure}

\paragraph{Calibration.} Inside the training distribution the response-level classifier is close to calibrated: ECE is 0.0317 on the prefixes over 30 split seeds (0.0216 on the stored text). On held-out conditions it is 0.0419, and across domains 0.096 (Section~\ref{sec:transfer-results}). The aggregate forest has ECE 0.116 inside the distribution, 0.1497 under leave-one-group-out and 0.4754 across domains. Figure~\ref{fig:calibration} shows the reliability curves. The calibration claim holds for use within one harness and one set of query domains.

\subsection{What Carries the Signal}\label{sec:signal}

Layout carries most of it. Table~\ref{tab:controls} gives the controls on the three text variants. On the stored text, length alone reaches 37.03\% with logistic regression and at most 38.79\% with nonlinear models. Cut to 800 characters, length falls to 31.72\%, the majority rate. The 24 formatting statistics reach 95.79\% on the prefixes against 97.84\% for the $n$-gram model. Masking brand spans and capitalised tokens changes the $n$-gram accuracy by less than 0.2 points in every variant.

\begin{table}[t]
\caption{What carries the response-level signal. Pooled five-system task, 5{,}028 responses (complete only: 3{,}921), mean over 30 split seeds of prompt-grouped 5-fold cross-validation; SD in parentheses. The formatting forest's seed varies with the split seed.}\label{tab:controls}
\centering\footnotesize
\begin{tabular}{@{}lrrr@{}}
\toprule
\textbf{Features} & \textbf{Stored text} & \textbf{800-char prefix} & \textbf{Complete only} \\
\midrule
Character $n$-grams & 98.14 (0.09) & 97.84 (0.21) & 97.86 (0.1) \\
Character $n$-grams, masked & 98.15 (0.07) & 97.72 (0.25) & 97.73 (0.06) \\
24 formatting statistics & 98.41 (0.09) & 95.79 (0.16) & 97.95 (0.25) \\
Length, logistic regression & 37.03 & 31.72 & 38.6 \\
Length, random forest & 36.93 & 31.72 & 41.56 \\
Length, gradient boosting & 38.79 & 31.55 & 41.84 \\
Majority class & 31.5 & 31.5 & 36.7 \\
\bottomrule
\end{tabular}
\end{table}

\begin{figure}[ht]
\centering
\includegraphics[width=0.95\textwidth]{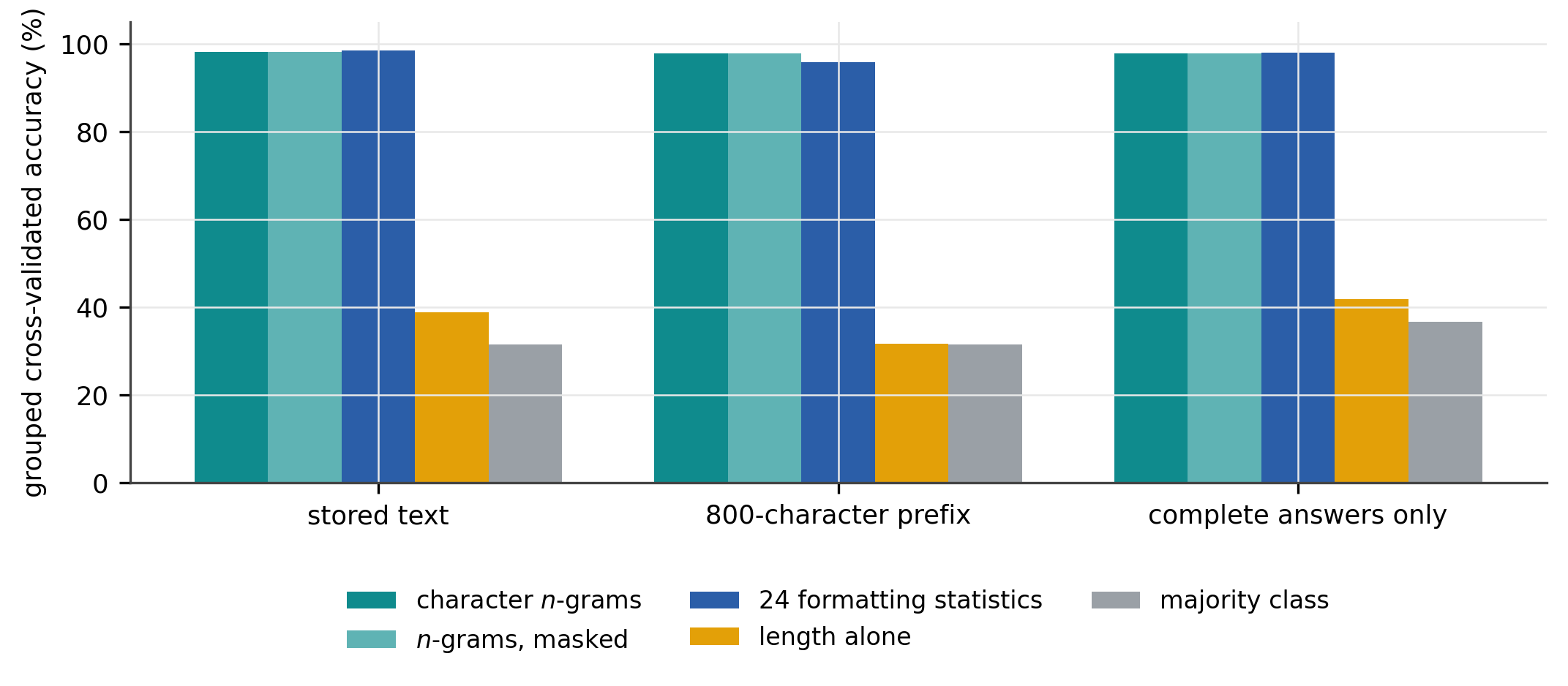}
\caption{Grouped cross-validated accuracy of the pooled five-system task by feature set, on the stored text, on 800-character prefixes and on complete responses only. Length is informative only while the text still shows where an answer stopped.}
\label{fig:signal}
\end{figure}

The leading formatting statistics on the stored text are the bracket-citation rate, mean sentence length and digit share. Part of what they read is the arm: the grounded reputation prompts ask for sources, and the completion caps differ between subsets for the same endpoint. The formatting control therefore measures the harness along with the system, and the response-level result is a statement about a deployed endpoint accessed through a fixed harness. Figure~\ref{fig:signal} shows the comparison.

\subsection{Empty Responses as a Sensitivity}\label{sec:empties}

Two policies test whether the 136 empty GPT-5.2 responses change a result. Dropping the 43 affected prompts for all three systems leaves the response-level category task at 99.97\% on the stored text (as built 99.97\%) and 100.0\% on prefixes. Treating an empty response as an observed answer gives 100.0\%, since only GPT-5.2 returns empty text. At the aggregate level, GPT-5.2's mean abstinence over the five category units is 0.0197 as built, 0.0125 with the affected prompts dropped and 0.1248 with empties counted as brand-free; its mean brand count is 3.891, 3.951 and 3.485, against 3.006 for Gemini~3 Flash under every policy. The GPT-5.2 abstinence figure depends on the policy; the ordering of brand volume does not.

\subsection{Aggregated Behavioural Profiles}\label{sec:unit-results}

At the unit the earlier version of this work classified, the twelve behavioural features separate four systems at 66.53\% (SD 3.23 over 30 split seeds, range 60.0 to 72.0; 66.47\%, SD 3.85, with the forest seed varied), group bootstrap 95\% interval [57.66, 76.61]. Leave-one-group-out gives 64.0\% and the RBF support vector machine 55.0\%. Macro F1 is 0.589.

Uniform chance, 25\%, is the wrong reference here. Grok units exist only in the gift domain and Perplexity units only in category ownership, so domain membership alone narrows every unit to three candidate labels. The within-group permutation null keeps that structure; its mean is 33.71\%, its 95th percentile 46.0\% and its maximum over 1{,}000 shuffles 60.0\%. The observed statistic at split seed 0, 64.0\%, gives $p = 0.001$. Pooled accuracy sits below both single-domain tasks: 71.31\% in the gift domain on 33 units (SD 6.30, leave-one-group-out 78.79\%) and 80.00\% in category ownership on 15 units. The category figure is identical across the 30 split seeds because five industry groups and five folds make grouped cross-validation the same as leave-one-group-out there; with the forest seed varied it is 73.11\% (SD 6.66).

Removing the three sentiment features gives 67.47\% (SD 2.92, leave-one-group-out 66.0\%, bootstrap [58.4, 76.4], permutation null 95th percentile 48.0\%). Other variations of the matrix stay inside a 3.3-point band: 67.53\% unweighted, 66.47\% with the earlier code's feature definitions, 64.93\% with 14 ambiguous single-word brand aliases removed, and 68.19\% on the 48 units left when the two parametric reputation units are dropped.

\paragraph{Grouped scores above ungrouped.} Under ungrouped 5-fold cross-validation the mean over 100 split seeds is 53.5\%, thirteen points below the grouped figure, and in category ownership 46.4\% against 80.0\%. Leakage would push the other way. The cause is visible in category ownership: a unit's nearest neighbour in the standardised feature space comes from the same industry 0.6 of the time (chance 0.143) and carries the same label only 0.133 of the time. Industry moves all three systems together more than the systems differ, so under ungrouped folds the same-industry units of the other systems sit in training and vote for the wrong label; grouped folds take them out. In the gift domain, where no unit's nearest neighbour shares its recipient condition, the two protocols differ by 3.4 points. The ungrouped figure is therefore evidence that condition effects dominate the aggregate features.

Per system, the classifier is uneven. F1 is 0.754 for GPT-5.2, 0.694 for Gemini, 0.604 for Grok and 0.305 for Perplexity, whose five units carry a recall of 0.253, with 67.3\% of them predicted as Gemini. A fifth of the label space rests on five units, and no statement about retrieval-augmented systems is made from them.

\begin{figure}[ht]
\centering
\includegraphics[width=0.92\textwidth]{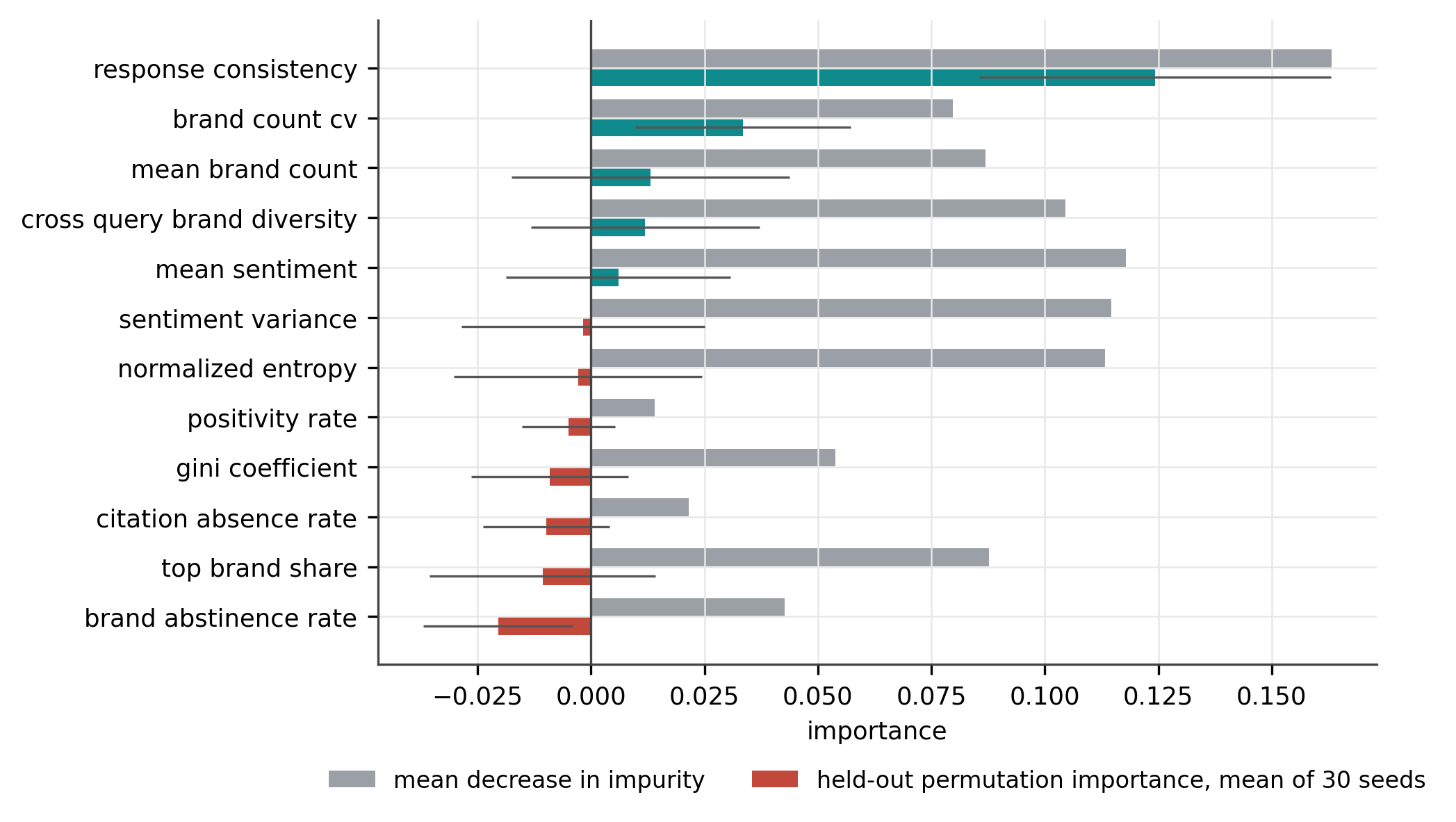}
\caption{Feature importance for the four-system aggregate classifier: mean decrease in impurity on the full fit, and permutation importance on held-out folds averaged over 30 split seeds with the forest seed varied (bars one SD).}
\label{fig:importance}
\end{figure}

\paragraph{What the forest uses.} Held-out permutation importance, averaged over the 30 split seeds with the forest seed varied, is largest for response consistency (0.1243, SD 0.0388, positive at every seed), followed by brand-count coefficient of variation (0.0335, SD 0.0237); 5 of the twelve have a positive mean, and brand abstinence has the most negative (\ensuremath{-}0.0205). On the 48 units without the parametric reputation subset, 7 features have a positive mean and the ranking below the first feature changes. Response consistency measures the harness as much as the system: its unit values run from 0.25 to 0.54 in the gift domain and from 0.62 to 0.75 in category ownership, it depends on the number of repetitions per prompt (40, 10, 5 or 1 by subset) and on sampling temperature, and it identifies Grok and Perplexity partly through the domain each is confined to. Part of the 66.53\% is system-domain confounding carried by this feature. Figure~\ref{fig:importance} shows the importances.

\begin{figure}[ht]
\centering
\includegraphics[width=0.92\textwidth]{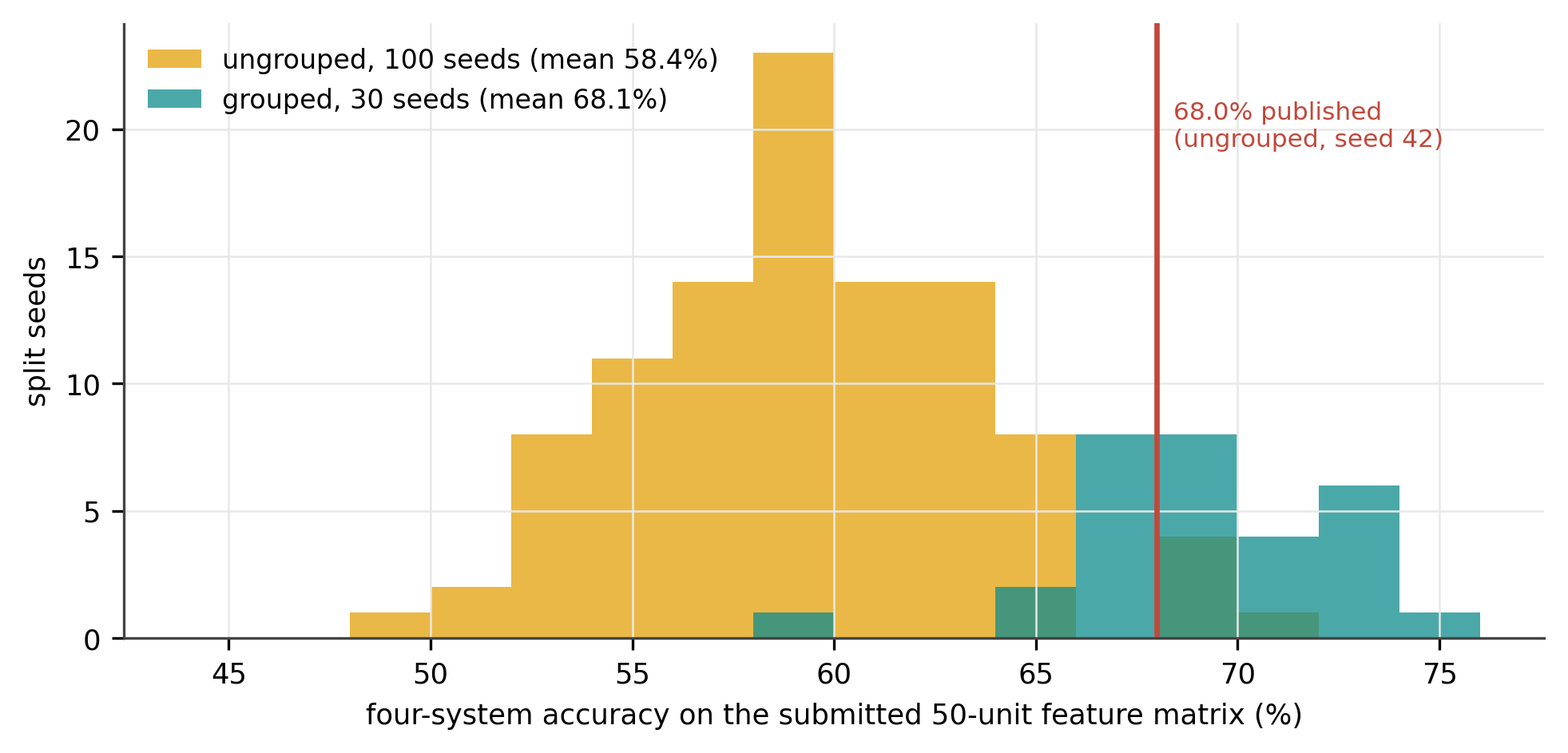}
\caption{Four-system accuracy over split seeds on the feature matrix the earlier version computed, under the ungrouped protocol that produced its headline (100 seeds) and the grouped protocol it described (30 seeds). The line marks the published 68.0\%, the ungrouped result at seed 42.}
\label{fig:seeds}
\end{figure}

\paragraph{The earlier headline.} The earlier version reported 68.0\% and described it as grouped. On the feature matrix that version computed, the 68.0\% is the ungrouped split at seed 42, which sits at the 99.0th percentile of 100 ungrouped seeds (mean 58.38\%). The grouped protocol on the same matrix gives 68.13\% over 30 seeds and 68.0\% under leave-one-group-out. The published value is close to what the described protocol gives, and it was produced by a different protocol. The fall to 66.53\% comes from rebuilding the matrix from the raw files (Appendix~\ref{secD}). Figure~\ref{fig:seeds} shows both distributions on the earlier matrix.

Principal component analysis on the standardised 50-unit matrix gives 35.66\%, 25.09\% and 15.77\% for the first three components. PC1 loads on brand volume against abstinence (mean brand count $+0.415$, brand-count CV $-0.399$, abstinence $-0.381$), PC2 on sentiment variance against sentiment level, PC3 on response consistency. The projection is descriptive.

\subsection{Cross-Domain Transfer at Both Units}\label{sec:transfer-results}

RQ3 is where the two units separate.

Table~\ref{tab:transfer} gives the four aggregate pairs. They are not independent trials: the ten category units are the test set of two pairs, so 44 test predictions cover 34 distinct units, and each pair is scored by a single fitted forest. The matched comparison with the response level is the gift and category pair in both directions, which scores 4 of 32 units. Its main component is one decision rule: a forest trained on the ten category units assigns all 22 gift units to the wrong system, with a mean stated confidence of 0.762. That is one sign reversal observed once, and the table reports it as counts. The pairs involving the parametric reputation subset rest on two units each.

\begin{table}[t]
\caption{Cross-domain transfer at the aggregate unit, Gemini~3 Flash against GPT-5.2, two classes. Features are the six brand-volume and concentration features defined in both domains; all twelve give the same four results. Each row is one forest fit.}\label{tab:transfer}
\centering\footnotesize
\begin{tabular}{@{}llrrrp{4.0cm}@{}}
\toprule
\textbf{Train} & \textbf{Test} & \textbf{Train units} & \textbf{Test units} & \textbf{Correct} & \textbf{Note} \\
\midrule
Gifts      & Category   & 22 & 10 &  4 & \\
Category   & Gifts      & 10 & 22 &  0 & every unit assigned to the other system \\
\midrule
Reputation & Category   &  2 & 10 &  5 & predicts one label for all ten \\
Category   & Reputation & 10 &  2 &  2 & two test units \\
\bottomrule
\end{tabular}
\end{table}

The reversal sits in the brand-volume features. Per response, Gemini~3 Flash names 8.41 unique brands against GPT-5.2's 0.94 in the gift domain, and 3.01 against 3.91 in category ownership. On complete answers only the figures are 8.84 (308 answers) against 0.95 (349) in gifts and 3.1 (211) against 3.91 (1{,}090) in category ownership; under the stricter completeness test, Gemini's category figure is 3.2 on 148 answers. In category ownership the reversal therefore rests on the minority of Gemini answers that finished. In the gift domain the dictionary undercounts GPT-5.2's brands, and the reference count of Section~\ref{sec:features}, 2.0 per answer, still leaves the gift ordering in place. The two domains also differ in brand vocabulary, completion cap and truncation rate, and none of these is ruled out, so the reversal is consistent with the transfer failure and has not been shown to cause it.

At the response level the same question has a different answer. Training on gifts and testing on category ownership and the reverse give a mean balanced accuracy of 94.06\% on the stored text, 89.92\% on 800-character prefixes, 97.63\% on complete answers and 93.07\% on masked prefixes (Appendix~\ref{secC}). These are two-class figures, GPT-5.2 against Gemini~3 Flash, and the two domains were collected between eight days and nine weeks apart under different scripts.

\begin{figure}[ht]
\centering
\includegraphics[width=0.95\textwidth]{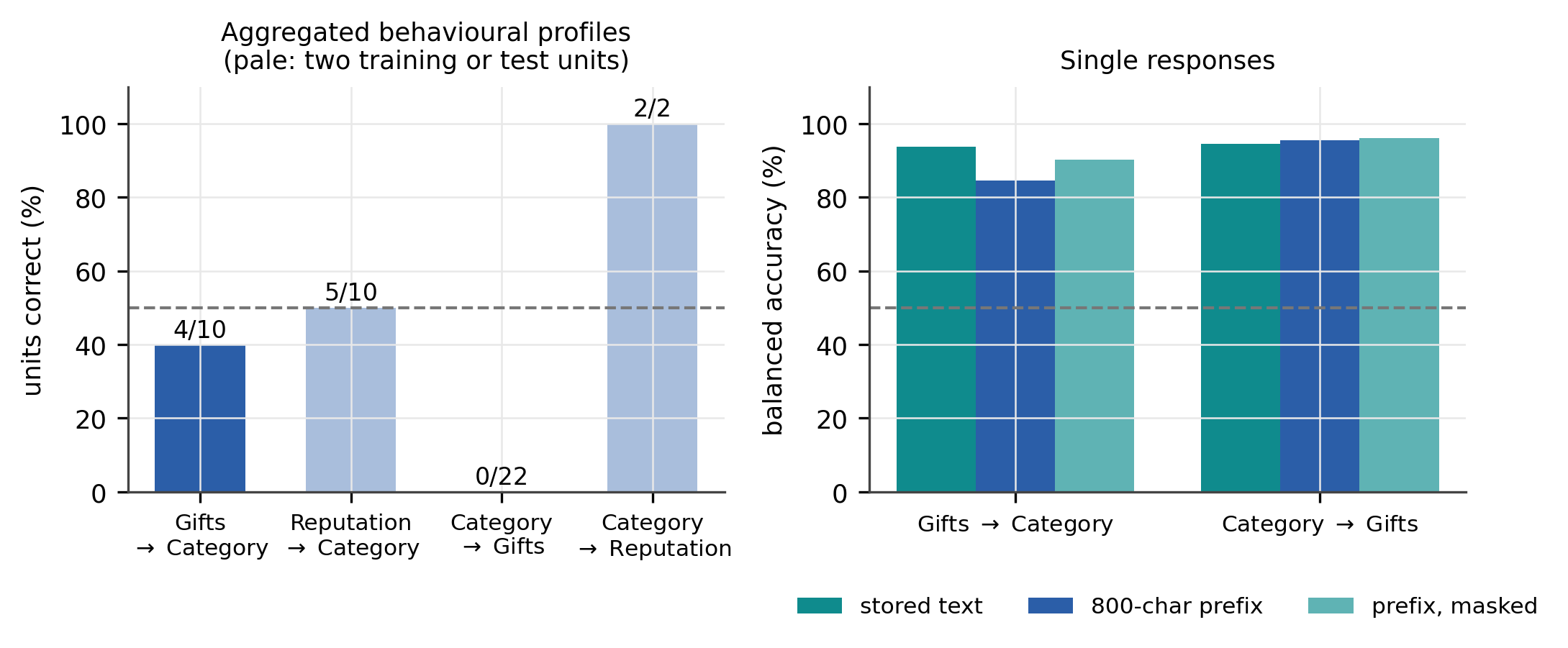}
\caption{The same cross-domain question at two units of observation, Gemini~3 Flash against GPT-5.2, chance 50\%. Left, aggregated behavioural profiles: units correct out of units tested for each of the four train and test pairs, one forest fit each. Right, single responses on the gift and category-ownership pairs, on the stored text, on 800-character prefixes and on masked prefixes.}
\label{fig:transfer}
\end{figure}

\subsection{The Grounded Arm}\label{sec:grounded}

The 360 responses of the grounded reputation arm are the only rows in this corpus that carry a structured source-attribution payload, and they are the rows the earlier pipeline never read. Within that arm, response-level attribution among Perplexity sonar-pro, Grok and Gemini~3 Flash with search is 99.86\% on the stored text and 99.91\% on prefixes.

The source-opacity feature the earlier version defined, the share of attributions typed as other or unattributable, is computable only here. Perplexity types 60.6\% of its 711 attributions that way, Grok 37.8\% of 994, and Gemini with search 76.0\% of 1{,}311, the last an upper bound because 939 of its domains were recovered from redirect titles. The parametric endpoints emit no attribution payload, so the feature is undefined for them. The earlier version reported roughly 5\% opacity for Perplexity against 70.2\% for the parametric endpoints; neither value is computable from these files, and the computable values order the other way.

The proportion of responses whose text contains no URL or bracket citation is 0.921 for Grok, 0.975 for Gemini with search, 0.992 for Gemini, 0.997 for Perplexity and 0.999 for GPT-5.2. A system that cites puts its sources in the structured payload, and its prose carries them no more often than any other system's.

\subsection{The Parametric Reputation Subset, Descriptively}\label{sec:reputation-descriptive}

The subset holds 1{,}296 responses about 24 companies from GPT-5.2 and Gemini~3 Flash. Its Gemini arm is truncated in storage, so every cross-system comparison drawn from it compares a complete answer with an opening clause. The quantity most affected is sentiment: the VADER gap between the two systems is $+0.511$ as stored (0.827 against 0.316) and $+0.043$ once GPT-5.2 text is cut to the paired Gemini length. Within this subset VADER has a Spearman correlation of 0.859 with stored response length. The subset is reported here and in the parametric row of Table~\ref{tab:response}, and nowhere else.

\subsection{Three-Way Agreement on the Leading Brand}\label{sec:agreement}

Of the 224 category-ownership queries in which all three systems named at least one brand, 42.4\% have the same most-mentioned brand across all three.

\section{Discussion}\label{sec:discussion}

\subsection{The Unit of Observation Decides the Answer}

One corpus, two units, two answers. A single response identifies its endpoint at 97.84\% over five systems with truncation removed, and GPT-5.2 against Gemini~3 Flash keeps 89.92\% balanced accuracy across a change of query domain. An aggregate of tens to hundreds of the same responses, summarised into brand volume, concentration, sentiment and source behaviour, separates four systems at 66.53\% inside the window and inverts across domains.

Aggregation discards the surface form, the part of the signal that crosses domains in this corpus, and keeps the brand behaviour, which moves with the domain. The brand-volume ordering of two systems flips between gift prompts and category-ownership prompts, and a profile built on that ordering flips with it.

A measurement that reports per-system brand behaviour reports at the aggregate unit. In this corpus a brand-behaviour profile measured in one query domain describes that domain, and carrying it to another domain needs evidence.

\subsection{Endpoint, Harness and Domain}

Section~\ref{sec:signal} shows that formatting statistics carry most of the response-level signal, and a formatting statistic reads the harness as readily as the endpoint. The grounded arm's request for sources is visible in the bracket-citation rate, completion caps differ between subsets, and one arm is truncated in storage.

The controls in this paper do not separate the endpoint from the harness. Within a subset every endpoint's harness is constant across conditions, and masking leaves the harness untouched, so held-out conditions and masking test generalisation over queries and over brand content. The cross-domain transfer changes the collection script and keeps 89.92\% balanced accuracy for two classes on prefixes. The one hold-out that changes the harness for three endpoints at once, the grounded arm, attributes every Perplexity answer correctly and no Grok answer correctly. The supported statement is about the pair: a deployed endpoint accessed through a fixed harness leaves a stable surface signature across the query conditions tested, and across the one harness change available, attribution holds for one endpoint and fails for another. Section~\ref{sec:followup} describes a collection that would separate the two.

\subsection{Three Applications, Bounded}

\textbf{Auditing a known collection.} An auditor who collects their own traffic through one harness can attribute answers to endpoints with a probability that is calibrated inside that harness and those query domains. Outside them the calibration measured here does not hold (ECE 0.096 across domains).

\textbf{Drift monitoring.} A per-endpoint surface signature that moves between two collections through the same harness is a flag for investigation. The present corpus covers one window and tests no drift.

\textbf{Reading AI-visibility numbers.} A brand-behaviour profile varies with the query domain, and in this corpus it does not travel between the two domains where it could be tested.

Content provenance for text collected in the wild is absent from this list by intent: the cross-harness evidence is one condition with a mixed result, and the surface signature is editable by anyone who wants to defeat it.

\subsection{Limitations}\label{sec:limitations}

\textbf{The design is not crossed.} No system appears in every domain. Grok appears in the gift and grounded subsets only; Perplexity in the category-ownership and grounded subsets only. The system label is partly determined by the domain.

\textbf{One collection window.} All data come from December 2025 to February 2026, with eight days to nine weeks between domains. Chen et al.~\cite{chen2024chatgpt} document behaviour shifts in one endpoint across months. Nothing here tests a second window.

\textbf{Truncation.} Three subsets contain cut answers. The 800-character prefix removes where an answer stops from the response-level task; it does not restore the content a cut answer lost, which the aggregate brand features of Gemini~3 Flash in category ownership are computed on.

\textbf{Aggregate sample size.} The aggregate classifier trains on 50 units in 17 groups, with 5 units for one of the four labels. Bootstrap intervals are wide by construction~\cite{varoquaux2018cross} and every aggregate-level accuracy carries a ten-point uncertainty.

\textbf{The sentiment features track length.} VADER's compound score has a Spearman correlation of 0.621 with response length over the corpus (0.859 in the parametric reputation subset), scores 94.1\% of responses positive and 76.3\% at or above 0.95, because a compound score summed over a long recommendation list saturates. No check against human coders exists, and no reliability coefficient is claimed. The three sentiment features act mainly as a length proxy, and the nine-feature result of Section~\ref{sec:unit-results} is the aggregate figure without them.

\textbf{The brand instrument.} Precision on a 50-answer sample is 0.827 and recall at most 0.43 (Section~\ref{sec:features}), checked by the analysis agent without a human rater. Fourteen single-word aliases are also common English nouns; removing them costs 1.6 points of aggregate accuracy.

\textbf{Thirteen prompts carry the gift domain.} Every response-level number for that domain rests on 13 prompt groups.

\textbf{Permutation resolution.} The response-level null uses 200 shuffles of a surrogate model, which bounds $p$ at 0.005.

\textbf{Grounded opacity is an upper bound.} Gemini with search returns its citations behind a redirect host, so 939 of its 1{,}311 stored attributions were typed from a domain recovered from the citation title.

\subsection{A Design That Would Settle the Confound}\label{sec:followup}

The collection that would settle the question has one harness, one completion cap large enough that no endpoint is cut, and one instruction set applied to every endpoint; logged version strings for every endpoint; at least two windows separated by three months; at least five query domains with at least 200 prompt templates each, so that every system appears in every domain; one open-weight model with frozen weights as an architectural control; and a second harness applied to the same endpoints and prompts, so that endpoint and harness are crossed. That design would split the surface signature into the part that belongs to the endpoint and the part that belongs to the harness.

\section{Conclusion}\label{sec:conclusion}

From a corpus of 6{,}324 analysable responses produced by five deployed endpoints in three query domains, a character $n$-gram classifier grouped by prompt attributes a single answer to its endpoint at 97.84\% on 5{,}028 responses cut to a common 800-character prefix (98.14\% on the stored text), keeps 97.43\% on held-out query conditions within the same harness (88.0\% unweighted, the small Valentine's conditions at 60.0 to 76.67\%), and is close to calibrated there. Twenty-four formatting statistics reach 95.79\% on the same prefixes. In the one held-out condition that changes the harness, Perplexity is attributed correctly and Grok is not.

The same corpus aggregated into 50 model-by-domain-by-condition units and summarised by twelve behavioural features separates four systems at 66.53\%, against a structure-preserving permutation null with 95th percentile 46.0\%, and one feature that tracks domain and repetition count carries most of that. Moved between the gift and category-ownership domains, the aggregate classifier gets 4 of 32 units right, while single responses keep 89.92\% balanced accuracy on the same two-system question.

The shape of an endpoint's answers carries its identity across the query domains tested. Its aggregated brand behaviour does not, and this design cannot say how much of that behaviour belongs to the system and how much to the domain or the harness. An AI-visibility measurement built on aggregated brand behaviour is bounded accordingly.

\begin{appendices}

\section{Per-System Detail, Aggregate Classifier}\label{secA}

\begin{table}[htbp]
\caption{Per-system precision, recall and F1 for the four-system aggregate random forest, pooled over 30 split seeds of stratified grouped 5-fold cross-validation. Support is aggregated units.}\label{tab:per_class}
\centering\footnotesize
\begin{tabular}{@{}lcccc@{}}
\toprule
\textbf{System} & \textbf{Precision} & \textbf{Recall} & \textbf{F1} & \textbf{Units} \\
\midrule
Gemini~3 Flash    & 0.643 & 0.753 & 0.694 & 17 \\
GPT-5.2           & 0.726 & 0.784 & 0.754 & 17 \\
Grok              & 0.696 & 0.533 & 0.604 & 11 \\
Perplexity (sys.) & 0.384 & 0.253 & 0.305 &  5 \\
\midrule
\textbf{Macro}    & \textbf{0.612} & \textbf{0.581} & \textbf{0.589} & \textbf{50} \\
\bottomrule
\end{tabular}
\end{table}

Row-normalised confusion, pooled over the same 30 seeds: 67.3\% of Perplexity units are predicted as Gemini, and 32.4\% of Grok units as GPT-5.2 against 8.8\% in the other direction.

\section{Per-System Feature Means}\label{secB}

\begin{table}[htbp]
\caption{Unit-averaged feature values by system, four-system matrix, 50 units, computed from the raw responses under the definitions of Section~\ref{sec:features}. Grok units are all gift units and Perplexity units all category-ownership units, so these columns mix system and domain.}\label{tab:profiles}
\centering\footnotesize
\begin{tabular}{@{}lcccc@{}}
\toprule
\textbf{Feature} & \textbf{Gemini~3 Flash} & \textbf{GPT-5.2} & \textbf{Grok} & \textbf{Perplexity (sys.)} \\
\midrule
Mean brand count $\mu_b$      & 6.801 & 1.797 & 5.304 & 2.897 \\
Brand-count CV                & 0.440 & 1.063 & 0.573 & 0.594 \\
Brand abstinence $\alpha$     & 0.055 & 0.292 & 0.130 & 0.082 \\
Gini $G$                      & 0.564 & 0.518 & 0.480 & 0.676 \\
$H_{\mathrm{norm}}$           & 0.818 & 0.662 & 0.833 & 0.722 \\
Top-brand share $\tau$        & 0.163 & 0.473 & 0.301 & 0.211 \\
Mean sentiment $\bar{s}$      & 0.948 & 0.981 & 0.998 & 0.967 \\
Sentiment variance            & 0.010 & 0.007 & 0.000 & 0.010 \\
Positivity rate $\pi$         & 0.970 & 0.997 & 1.000 & 0.997 \\
Response consistency $\bar{c}$ & 0.504 & 0.524 & 0.400 & 0.644 \\
Citation-absence rate         & 0.992 & 0.999 & 0.997 & 0.998 \\
Cross-query diversity $D_q$   & 21.82 & 4.379 & 23.85 & 0.564 \\
\bottomrule
\end{tabular}
\end{table}

\section{Response-Level Transfer, Per Pair}\label{secC}

\begin{table}[htbp]
\caption{Response-level cross-domain transfer, Gemini~3 Flash against GPT-5.2, balanced accuracy (\%) and expected calibration error. One fit per pair and variant. Masked: dictionary brand spans, capitalised tokens, URLs and digits replaced.}\label{tab:transfer_text}
\centering\footnotesize
\begin{tabular}{@{}lrrrrr@{}}
\toprule
\textbf{Train $\rightarrow$ test} & \textbf{Stored} & \textbf{Prefix} & \textbf{Complete} & \textbf{Masked prefix} & \textbf{ECE, prefix} \\
\midrule
Gifts $\rightarrow$ Category & 93.67 & 84.43 & 96.24 & 90.17 & 0.1156 \\
Category $\rightarrow$ Gifts & 94.44 & 95.41 & 99.03 & 95.97 & 0.0763 \\
\midrule
\textbf{Mean} & \textbf{94.06} & \textbf{89.92} & \textbf{97.63} & \textbf{93.07} & \textbf{0.096} \\
\bottomrule
\end{tabular}
\end{table}

Pairs that train or test on the parametric reputation subset are not reported, because its Gemini arm is truncated in storage.

\section{Corrections to Earlier Reported Values}\label{secD}

This appendix documents the difference between the present manuscript and an earlier, unpublished version of this analysis, some of whose values have appeared in talks, posts and a book, so that anyone who has cited the earlier values can correct the citation. Every corrected value comes from a rerun of the analysis against the raw response files.

\subsection{Why the Numbers Changed}

Five defects in the earlier pipeline account for most of the differences.

A type gate in the loader silently skipped 72 files, so 360 responses of the grounded reputation arm never entered any analysis, and the corpus total reported in the earlier text is not reachable from the files.

136 empty responses were counted as answers naming no brand, which inflated one system's brand-abstinence rate and depressed its mean brand count.

The headline accuracy of 68.0\% came from an ungrouped 5-fold split at seed 42, and the text described stratified grouped $K$-fold. On the earlier feature matrix that seed is at the 99th percentile of 100 ungrouped seeds (mean 58.38\%), and the grouped protocol gives 68.13\%. The published value sits close to the grouped mean on its own matrix; the corrected 66.53\% differs because the matrix was rebuilt.

The sentiment instrument that ran was TextBlob on the first 200 responses of a unit, and the text named VADER. The grouped cross-validation, permutation test, bootstrap intervals and cross-window replication the text described have no code that produced the reported values.

Several values in the feature-profile tables were never computed by this pipeline: some come from earlier manuscripts on other corpora, and some from a subset the pipeline treats differently.

\subsection{Values That Changed}

{\scriptsize\setlength{\tabcolsep}{3pt}
\begin{longtable}{@{}>{\raggedright\arraybackslash}p{2.9cm}>{\raggedright\arraybackslash}p{2.4cm}>{\raggedright\arraybackslash}p{3.0cm}>{\raggedright\arraybackslash}p{3.3cm}@{}}
\caption{Earlier value against corrected value. Every corrected value is produced by the rerun described in Section~\ref{sec:methods}.}\label{tab:changes}\\
\toprule
\textbf{Quantity} & \textbf{Earlier} & \textbf{Corrected} & \textbf{Reason} \\
\midrule
\endfirsthead
\toprule
\textbf{Quantity} & \textbf{Earlier} & \textbf{Corrected} & \textbf{Reason} \\
\midrule
\endhead
\bottomrule
\endlastfoot
Raw corpus size & 6{,}340 responses & 6{,}475 on disk, 6{,}324 analysable & the loader reached 6{,}115 rows; 6{,}340 is not reachable from the files \\
Gift subset, raw $n$ & 1{,}279 & 1{,}054 & every file present was read; 1{,}279 exceeds the design ceiling \\
Category subset, raw $n$ & 3{,}750 & 3{,}750 stored, 3{,}614 analysable & 136 empty responses are missing data \\
Grounded reputation arm & not reported & 360 responses in 72 files & recovered from the loader gate \\
Aggregated units per subset & 30 / 5 / 15 & 33 / 2 / 15 & recomputed; the Reputation row also listed a system with no rows there \\
Headline four-system accuracy & 68.0\% & 66.53\% (SD 3.23, 30 grouped seeds) & the published value was an ungrouped split at one seed; grouped on the earlier matrix gives 68.13\% \\
Chance reference & 25\% & permutation null mean 33.71\%, 95th pct.\ 46.0\% & two of four labels exist in one domain only \\
Bootstrap CI on the headline & [55, 80] & [57.66, 76.61] & 1{,}000 resamples at the group level \\
Leave-one-group-out & 58.0\% & 64.0\% & the published figure was leave-one-\emph{unit}-out \\
SVM baseline & 56.0\% & 55.0\% & same protocol as the forest \\
Macro F1, four systems & 0.63 & 0.589 & pooled over 30 grouped seeds \\
Gift-domain accuracy & 63.6\% on 30 units & 71.31\% on 33 units & 33 units exist in that subset \\
Category-domain accuracy & 80.0\%, $p = 0.004$ & 80.00\%, $p = 0.001$ & accuracy reproduces; the null is now 1{,}000 within-group shuffles \\
Class weighting & described as class-balanced & applied; unweighted gives 67.53\% & the earlier code described weighting it did not apply \\
PCA variance, first three & 32.9 / 26.0 / 14.0 & 35.66 / 25.09 / 15.77 & the matrix behind it changed \\
Top features by importance & response consistency, brand-count CV, mean brand count & response consistency, brand-count CV; 5 of 12 positive & the published figure labelled in-sample values as held-out \\
Mean brand count by system (Study~1) & 8.66 / 0.59 / 7.37 & 8.41 / 0.94 / 7.69 per response, gift domain & the published values are another manuscript's table \\
Brand abstinence by system & 0\% / 70\% / 0\% / 0\% & 0.055 / 0.292 / 0.130 / 0.082 & computed over non-empty responses \\
Study~2 mean sentiment & 0.83 / 0.32 & 0.827 / 0.316 as stored; gap $+0.043$ length-matched & reproduces as a number; invalid because of truncation \\
Study~2 positivity rate & 96.4\% / 51.5\% & 97.4\% / 51.1\% as stored & reproduces as a number; invalid because of truncation \\
Response consistency & 0.51 / 0.57 & 0.504 / 0.524 & computed per prompt across its repetitions, as defined \\
Transfer accuracy by pair & 40.0 / 50.0 / 77.3 / 50.0, mean 54.3\% & 4/10, 5/10, 0/22, 2/2 & the published 77.3\% came from the earlier feature values; reported as counts, one fit each \\
Transfer degradation & $-17.5$ pp & not reported & the published column compared three different baselines, and one pair has two test units \\
Repetitions per prompt & five for every prompt & 40 / 10 / 5 / 1 by subset & from the stored rows \\
System version strings & GPT-5.2, Gemini~3 Flash, Grok-4-1, sonar-pro & Grok version not logged; the others as recorded & nothing on disk supports the Grok version string \\
Collection window & 2025-12-04 to 2026-02-28 & 2025-12-23 to 2026-02-24 & per-row timestamps \\
Three-way brand agreement & 41.6\% & 42.41\% over 224 queries & reproduces \\
Response-level attribution & not reported & 97.84\% on 5{,}028 responses, 800-character prefixes & the result the corpus supports \\
\end{longtable}}

\subsection{Claims Withdrawn}

The following claims of the earlier version are withdrawn.

\textbf{The sentiment reliability coefficient.} ``Cohen's $\kappa = 0.72$ between VADER and the judge model'' on a stratified sample of 200 responses. No code, sample or result file behind it exists, and the instrument the pipeline ran was TextBlob. No agreement against human coders has been measured.

\textbf{The cross-window replication.} The section reporting that an independent April 2026 multilingual corpus reproduces the GPT-versus-Gemini sentiment gap at $+0.18$, with GPT higher for 58 of 66 brands. No code or result file stands behind those values. That corpus covers different model versions (GPT-5.4 and Gemini~3.1 Pro), so it could not replicate the same systems. Under the instrument it was scored with, a multilingual transformer, it gives $+0.0895$ with GPT higher for 65 of 66 brands. VADER is an English lexicon: on all twelve languages it gives $+0.0525$ (51 of 66), a value without meaning for non-English text, and on the English rows $\ensuremath{-}0.093$ (3 of 66). The February gap the section set out to replicate falls from $+0.511$ to $+0.043$ under its own truncation control.

\textbf{The Study 2 sentiment gap and the profile rows built on it.} $\bar{s} = 0.83$ against $0.32$ and positivity 96.4\% against 51.5\%. These reproduce from the stored rows under VADER (0.827 against 0.316; 97.4\% against 51.1\%), and they are invalid because the Gemini arm is truncated in storage: with lengths matched the gap is $+0.043$. The Study~1 brand-volume row of the same table (8.66 / 0.59 / 7.37) is imported from a different manuscript on recipient-split cells.

\textbf{Source opacity.} 70.2\% for the parametric endpoints and roughly 5\% for Perplexity, and the foundation-versus-RAG section built on them. Neither value is computable from these files, and the computable values order the other way (Section~\ref{sec:grounded}).

\textbf{Per-system transfer.} The appendix table reporting 73.6\% cross-domain accuracy for GPT-5.2 and 35.0\% for Gemini. Two of the four transfer pairs have two training or two test units and one predicts a single label, so a per-system split of four pairs is not interpretable.

\textbf{The missingness-indicator design.} The description of a 24-column matrix pairing each feature with a missingness indicator. The four-system matrix has no missing cells, so every indicator column would be constant.

\textbf{The feature-importance agreement claim.} ``MDI and permutation importance agree on the top three.'' They disagree (Section~\ref{sec:unit-results}).

\textbf{Three collection-protocol statements.} Five repetitions for every prompt, the Grok version string, and the collection window endpoints. The stored rows contradict all three.

\textbf{Both availability statements.} ``Available from the corresponding author upon reasonable request'' for the data and for the code. They are replaced by the statements under Data Availability and Code Availability.

\subsection{What Holds}

Five quantities of the earlier version reproduce: the raw $n$ of the parametric reputation subset (1{,}311 stored rows), the 50-unit aggregate total, the category-ownership accuracy of 80.0\%, the three-way agreement on the leading brand (41.6\% published, 42.41\% rebuilt), and, on the earlier feature matrix, the 68.0\% headline as a value that the described grouped protocol also gives. The permutation $p$ of 0.001 now reproduces, from a test the earlier code did not run.

\end{appendices}

\backmatter
\bookmarksetup{startatroot}

\bmhead{Data Availability}

Of the 6{,}475 stored responses, 574 in the gift subset are byte-identical to files in \url{https://doi.org/10.5281/zenodo.21904788} (CC BY 4.0). The 3{,}750 category-ownership responses were collected for the author's Category Ownership Map dataset; the response file analysed here has md5 digest \texttt{90f9edc1a8613c056f580ebe21455715}, against which any copy of that dataset can be checked. The remaining 480 gift responses, the per-response table used in every analysis, the 50-unit feature matrix, the per-file manifest with md5 digests, the 100-response sentiment sample labelled by a single automated pass against a written codebook with no human rater, and the 50-answer brand-dictionary check are available from the author on request. They will be deposited on Zenodo under CC BY 4.0, and the deposit DOI will be added to a later version of this paper. The 1{,}311 parametric and 360 grounded reputation responses are not released and are not available on request: the reputation prompts name 24 companies, some of which are clients of the author's company, and no response about a client is published. The per-response table and the feature matrix are released without the text of those responses.

\bmhead{Code Availability}

The analysis pipeline, from the raw response files to every number and figure in this paper, is a set of numbered Python scripts, each of which writes the JSON result file that the manuscript's numbers are filled from. It is available from the author on request and will be deposited with the data. The data-collection infrastructure for the source corpora is the open-source Dice Roller package, \url{https://github.com/Rankfor/rankfor-open/tree/main/dice-roller}.

\bmhead{Acknowledgements}

The author thanks the Estonian Entrepreneurship University of Applied Sciences for institutional support.

\section*{Declarations}

\begin{itemize}
\item \textbf{Funding:} This research received no external funding. Rankfor.AI paid for the API access used to collect the corpus.
\item \textbf{Competing interests:} The author declares a commercial interest. D.~\.Zatuchin founded Rankfor.AI, which sells AI brand-visibility measurement, and the attribution methods studied here are relevant to that business. The corpora were collected with Rankfor.AI's Dice Roller software for studies run by Rankfor.AI, and Rankfor.AI is one of the 24 companies named in the reputation prompts. Section~\ref{sec:discussion} states a boundary on what measurements of the kind that company sells can claim.
\item \textbf{Ethics approval:} Not applicable. The study analysed publicly accessible AI systems and did not involve human subjects.
\item \textbf{Consent to participate:} Not applicable.
\item \textbf{Consent for publication:} Not applicable.
\item \textbf{Use of AI tools:} The author used an AI assistant (Claude, Anthropic) for grammar and language editing, and to draw the figures from the result files by script. The 100-response sentiment sample and the 50-answer brand-dictionary check were labelled by an LLM agent against a written codebook, with no human rater, as stated under Data Availability. The author reviewed all of this output and takes full responsibility for the content.
\item \textbf{Author contribution:} D.\.Z.\ designed the analysis, collected the source corpora, performed the classification experiments and wrote the manuscript.
\end{itemize}

\bibliography{references}

\end{document}